\documentclass[12pt]{article}
\begin{document}
\baselineskip .3in
\begin{titlepage}
\begin{center}
\vskip .2in {\bf A. Chandra}.{\bf  A. Bhattacharya }.{\bf B.
Chakrabarti }

\vskip .2in{\large{\bf Strangeness in proton and  properties of
nucleons in nuclear matter revisited }}\vskip .2in
\end{center}
\vskip .1in

\vskip .3in {\bf Abstract} The properties of the nucleons in
nuclear medium have been investigated in the context of the flux
tube model incorporating strangeness $(s\overline{s})$
contribution to proton structure in conformity with the
experimental indication. Proton is described as a pentaquark
system with strange quark contribution whereas neutron is
described in three quark configuration. The Quasi particle model
of diquark is used to describe the structures of the nucleons.
Modifications of the properties like swelling, mass,
incompressibility, ratio of the structure functions
($\frac{F_{2}^{n}(x)}{F_{2}^{p}(x)}$), Gottfried Sum rule for
nucleons in nuclear medium have been studied and significant
effects have been observed. It has been suggested that the change
of the size degree of freedom of the nucleon in the nuclear medium
plays an important role in describing the properties in medium.
The results are discussed in detail and compared with existing
experimental and theoretical predictions. Some interesting
observations are made.
 \vskip .2in
 {\bf keywords} Structure Function.Gottfried Sum Rule.Diquark.Quasi Particle


\vskip .2in A.Chandra

Department of Physics, Jadavpur University, Calcutta 700032, India.\\
arpita1chandra@gmail.com

A.Bhattacharya

Department of Physics, Jadavpur University, Calcutta 700032, India.\\
pampa@phys.jdvu.ac.in

B.Chakrabarti

Department of Physics, Jogamaya Devi College, Kolkata, India\\
ballari$_{-}$chakrabarti@yahoo.co.in

  \vskip .2in

\newpage
\vskip 0.4in
\end{titlepage}

\newpage
 \vskip.25in

\vskip .3in \noindent {\bf 1. Introduction}

The modification of the internal structure of nucleons in the
nuclear environment was studied by European Muon Collaboration
(EMC)[1]. The experiment predicted the swelling of the nucleon in
nuclear matter and has observed that the properties of the nucleon
in nuclear matter differ substantially from that of a free
nucleon. The swelling is largely interpreted as the effect of
attraction of  neighbouring quarks to the quarks constituting the
nucleon. Recently Seely et al [2] have reported that the
modification of the particle properties primarily depend on the
immediate neighborhood within nucleus and not on the mass or
density of the nucleus. Wu et al [3] have studied the modification
of the properties of the nucleon in nuclear matter and in finite
nuclei describing the nucleon as non-topological solitons. They
have considered that the quarks inside the soliton bag not only
couple to the scalar field that binds the quarks together into
nucleons but also to the additional meson field generated by
nuclear environment. Kim et al.[4] have studied the swollen of
nucleon in nuclear matter for (e,e'p)reaction. They have
incorporated the effect to the form factors containing the
anomalous magnetic moment of the nucleons and studied the effect
by exchanging the form factors of the free nucleons by those of
the nucleons in nuclear matter. Recently Rozynek [5] has studied
the modification of the nucleon structure function in nuclear
matter in the context of relativistic mean field(RMF) approach.
Higinbotham et al [6] have investigated the nuclear short range
correlation (SRC) and the EMC Effect using the observed
phenomenological relationships. They have suggested that the
correlation between the EMC and the SRC is dominated by the high
momentum nucleons in the nucleus.

   Recent experiments have suggested [7] a strange contribution
   to the proton magnetic moment. It has been observed that
 the fraction of the spin of the proton carried by quarks is
small and a substantial contribution comes from the strange
quark-antiquarks $(s\overline{s}$. A number of works have been
done on the strangeness contribution to the proton structure. Zou
et al [8] have investigated the strangeness contribution to the
proton magnetic moment considering the proton as a
[uuds$\overline{s}$] system. Riska et al [9] have investigated the
magnetic form factor of the proton in uuds$\overline{s}$
configuration. They have pointed out that empirical strange form
factor may be described with the 15 percent admixture of
uuds$\overline{s}$. They have also observed that the transition
matrix element between uud and uuds$\overline{s}$ components gives
significant contribution. An et al [10] have studied
uuds$\overline{s}$ configuration of proton and have investigated
the strangeness magnetic moment considering one of the constituent
quarks in excited state. They have observed a significant
contribution in qqqq$\overline{q}$ in baryon and have revealed the
possibility of the colored cluster configuration (diquarks) rather
than the 'meson cloud' configuration.

In the present work we have studied the modification of the
properties of the nucleons in nuclear media incorporating a
strange quark contribution to the proton structure. The proton has
been considered as a five quark system with a $s\bar{s}$
contribution and supposed to have a pentaquark configuration like
[uuds$\bar{s}$] such as $[ud]_{0}[us]_{0}\bar{s}$ where [ud] and
[ud] are scalar diquarks of corresponding flavours. We have
treated intrinsic sea contribution s$\overline{s}$ valence like.
In this context it may be mentioned that recently An et al [10]
have pointed out that the contribution from intrinsic sea is much
more valance like particularly at large x- region. The neutron has
been considered in usual diquark-quark system [$(ud)_{0}d]$. It
would be interesting to investigate the fact that how the
incorporation of strange quark degrees of freedom to the proton
and diquark-diquark-anti quark configuration reproduces the
properties of the proton in nuclear medium. A comparison between
the properties of the neutron and the proton in medium have been
studied. The quasi particle model for diquark [11] has been used
to estimate effective mass of diquarks. The effect of the nuclear
medium on the properties like the radius, the compressibility,
Roper resonance of nucleons have been investigated through the
change of the radii of the nucleons in nuclear medium. The
variation of the $\frac{F_{2}^{n}(x)}{F_{2}^{p}(x)}$ and the
Gottfried sum rule in nuclear medium have been studied. So far
most of the studies of the structure functions of nucleons are
done for the free nucleons. It would be interesting to investigate
how the swollen of the nucleons in nuclear medium affects the
structure functions. It is interesting to observe in the current
investigation that the pentaquark configuration of the proton is
consistent with the $\overline{u}$-$\overline{d}$ asymmetry in the
nucleon sea distribution.

 \vskip .25in\vskip .3in \noindent {\bf 2. The Model}

 Recently [11] we have suggested a model in which two quarks are
assumed to be correlated to form a low energy configuration, a
diquark. Diquarks are supposed to behave like a  quasi particle in
an analogy with an electron in the crystal lattice which behaves
as a quasi particle [12]. It is well known that a quasi particle
is a low-lying excited state whose motion is modified by the
interactions within the system. An electron in a crystal is
subjected to two types of forces, namely, the effect of the
crystal field ($\nabla V$) and an external force (F) which
accelerates the electron [12]. Under the influence of these two
forces, an electron in a crystal behaves like a quasi particle
having velocity $v$ whose effective mass $m^{*}$ reflects the
inertia of electrons which are already in a crystal field. The
effective mass can be represented as:
\begin {equation}
m^{*}\frac{dv}{dt}= F
\end{equation}
 Again the bare electrons ( with normal mass) are affected by the lattice
force -$\nabla V$(where V is periodic potential) and the external
force F so that:
\begin{equation}
m\frac{dv}{dt}= F - \frac{dV}{dx}
\end{equation}
 So the ratio of the normal mass (m) to the effective mass ($m^*$) can be
expressed as:
 \begin{equation}
 m/m^{*}= 1 - \frac{1}{F}[\frac{\delta{V}}{\delta x}]
 \end{equation}
An elementary particle in vacuum may be suggested to be in a
situation exactly resembling that of an electron in a crystal
[12]. We have proposed a similar type of picture for the diquark
$[ud]_{0}$ as a quasi particle inside a nucleon. The strong
interaction is characterized by two types of forces. One is short
range confinement force and the other one is the asymptotic
freedom. We have assumed that the formation of diquark is favoured
by the nature of the vacuum and the two types of forces within the
hadrons. The diquark is supposed to behave like an independent
hypothetical colour antisymmetric body under the influence of
these two types of forces. One is represented by the potential V
=$-\frac{2}{3}$$\frac{\alpha_{s}}{r}$ where $\alpha_{s}$ is the
strong coupling constant and this potential is assumed to resemble
the crystal field on a crystal electron. On the other hand we have
considered an average force F = -ar for the external force where
'a' is a suitable constant. It has been assumed that under the
influence of these two types of interactions the diquark behaves
like a quasi particle, a low lying excited state and its mass gets
modified. The potential can be represented as:

\begin{equation}
V_{ij}=-\frac{\alpha}{r}+(F_{i}.F_{j})(-\frac{1}{2}Kr^{2})
\end{equation}

Where the coupling constant
$\alpha$=(2/3)$\alpha_{s}$,$F_{i}$.$F_{j}$=-(2/3), K is the
strength parameter. Hence $V_{ij}$ may be represented as:

\begin{equation}
V_{ij}=-\frac{(2/3)\alpha_{s}}{r}+ar^{2}
\end{equation}
Where a=K/3.

The ratio of the constituent mass and the effective mass of the
diquark (m$_{D}$) has been obtained by using the same formalism as
in equation (3) and we arrive at:
 \begin{equation}
 \frac{m_{q}+m_{q\prime}}{m_{D}}= 1+ \frac{\alpha_{s}}{3ar^{3}}
 \end{equation}

Here $m_{q}$+m$_{q\prime}$ represents the normal constituent
masses of the quarks forming the relevant diquarks and $m_{D}$ is
the effective mass of the diquark.
$\alpha$=$\frac{2}{3}\alpha_{s}$, $\alpha_{s}=0.58$ [13] and the
strength parameter a = 0.02$GeV^{3}$, 'r' is the radius parameter
of the diquark. The radii parameter of the scalar diquarks have
been used from existing literature as $r_{ud}$ =0.717 fm[14],
$r_{us}$=0.616 fm[14]. Using the masses of the constituent quarks
($m_{u}$=$m_{d}$= 0.360 GeV and $m_{s}$=0.540GeV )[15], we have
computed the masses of the [$ud_{0}$] and the [$us_{0}$] diquarks
as $m_{ud}=0.590 GeV$, $m_{us}=0.674 GeV$ respectively. These
diquark masses have been used to investigate the properties of the
proton and the neutron in subsequent analysis. \vskip .25in

\vskip .3in \noindent {\bf 3. Formulation}

 Mathieu et al.[16] have considered a flux tube model of the
 nucleons with quark-diquark configuration where the quark and the diquark are
linked by the flux
 tube. The hamiltonian(H) for an isolated nucleon has been represented as,

 \begin{equation}
H =\frac{p^{2}}{2\mu}+\sigma r-\frac{4\alpha_{s}}{3r}
\end{equation}

Where '$\mu$' is the reduced mass of the system, 'p' is the
relative momentum, $\sigma$ is the string tension and $\alpha_{s}$
is the strong coupling constant. It is known that when the nucleon
is in the nuclear medium the hamiltonian changes due to the effect
of the neighbouring nucleons. In the presence of other nucleon the
flux tube are topologically arranged in such a way that the linear
potential of the system is minimized. When the constituents of two
nucleons are very close to each other, the flux tubes are
suggested to be redistributed among the constituents to yield a
topology where a lowering in total length for the flux tubes
occurs. Considering the effect of one perturbing nucleon, the
effective modified linear potential [16] may be represented as ,

\begin{equation}
V_{L eff}(r) = \sigma r- \frac{\sigma
\rho}{6}\frac{\pi}{24}\int|\psi(r')|^{2}(t^{2}-4u^{2})^{2}\theta(t-2u)d^{3}r'
\end{equation}

The one gluon exchanged term would also be modified partially due
 to the presence of the other nucleons and the effective modified coulomb potential may be represented as in [16]:
 \begin{equation}
 V_{C eff}(r) = -\frac{4\alpha_{s}}{3r}-\frac{4 \alpha_{s} \rho}{18}(\frac{2\pi}{3})\int|\psi(r')|^{2}
 (1-z')(r^{2}-r'^{2}-4rr')d^{3}r'
 \end{equation}

Where 'r' and '$r'$' represents the length of the flux tube for
two nucleon clustering as in [16]. $\Psi(r')$ is the nucleon wave
function, $\rho$ is the nuclear matter density and $\sigma$ is the
string constant. Symbols are used as in Mathieu et al [16].
 The total effective potential($<V>$)
between the constituents of a nucleon can be expressed as:

  \begin{equation}
 <V> = \int V_{eff}(r)| \psi(r)|^{2}d^{3}r
 \end{equation}
 where,$ V_{eff}(r)$=$V_{L eff}(r)$ + V$_{C eff}(r)$ and
 $\Psi(r)$ is the wave function of a typical nucleon. Thus the
hamiltonian can be expressed as:

\begin{equation}
< H >= <\frac{p^{2}}{2m}> + <V>
 \end{equation}

To estimate the energy we need the wave function for the nucleons.
We have used the wave function for the nucleons suggested by the
Statistical model [17]. The wave function in the ground state runs
as[17],
\begin{equation}
|\Psi(r)|^{2}=\frac{315}{64\pi r_{0}^{9/2}}(r_{0}-r)^{3/2}\theta
(r_{0}-r)
\end{equation}
Where $r_{0}$ is the radius parameter of the corresponding hadron
and $\theta(r_{0}-r)$ represents the step function. We have used
this wave function for the nucleons in our subsequent studies of
the properties of the nucleon in nuclear matter. It may be
mentioned that Mathieu et al [16] have used the Gaussian type of
wave function for the nucleons in their investigation.

Considering proton as a five quark system with
diquark-diquark-antiquark configuration as stated earlier, the
Hamiltonian for the proton in reduced mass system is obtained as:

\begin{equation}
<H_{p}>=\frac{29.678}{r_{p}^{2}}+0.545(\sigma
r_{p})-0.055(\sigma\rho
r_{p}^{4})-\frac{3\alpha_{s}}{r_{p}}+0.312(\alpha_{s}\rho
r_{p}^{2})
\end{equation}

Similarly the Hamiltonian for the neutron in quark-diquark
configuration has been estimated and is expressed as:

\begin{equation}
<H_{n}>=\frac{26.366}{r_{n}^{2}}+0.545(\sigma
r_{n})-0.055(\sigma\rho
r_{n}^{4})-\frac{3\alpha_{s}}{r_{n}}+0.312(\alpha_{s}\rho
r_{n}^{2})
\end{equation}

Minimizing the Hamiltonian in (13) and (14) with respect to the
radius parameter we come across the following expressions:
\begin{equation}
\frac{r_{p}}{r_{p}^{0}}=1.259 [1 + \sqrt{(1-175.73 \rho)} ]
^{\frac{-1}{3}}
\end{equation}

\begin{equation}
\frac{r_{n}}{r_{n}^{0}}=1.260 [1 + \sqrt{(1-156.618 \rho)} ]
^{\frac{-1}{3}}
\end{equation}
 Where ${r_{p}^{0}}, {r_{n}^{0}}$ are the radius of the proton and the neutron respectively at $\rho$=0, $\rho$
 is the density of the nuclear medium, $r_{p}$, $r_{n}$ are the radius of proton and the neutron respectively in
 nuclear medium. The string tension $\sigma$ is taken as 0.999$GeV^{2}$ form [18].
 GeV$^2$. We have neglected the coulomb term to arrive at the expressions
 (15) and (16).

  We have studied the variation of '$r/r^{0}$' with matter density $\rho$ and it has been
 displayed in the Figure-1. The critical density is defined as the density at which the nucleon
 ceases to exist and in the present formulation it represents the density at which
 the radius of the nucleons in the nuclear medium becomes imaginary. We have estimated the the critical densities for the
 for proton and the
 neutron as 4.17 and 4.69 times the normal
 nuclear density respectively. The ratio of the masses of the nucleons in nuclear medium '$M^{*}$' to that of the free
 nucleon '$M_{0}$' have been estimated for both the proton and the neutron and have displayed in the Figure-2.

For a bound nucleon the expression for the incompressibility runs
as,
\begin{equation}
K=\frac{1}{3}r^{2}\frac{d^{2}E}{dr^{2}}
\end{equation}
  We have estimated $\Delta K$, the decrease
in the incompressibility from free to normal nuclear medium for
the proton and the neutron as 0.156 GeV and 0.132 GeV
respectively. The variation of $\frac{K}{K_{0}}$ with $\rho$ have
been shown in Figure-3 where $K_{0}$ is the incompressibility at
$\rho$=0.

The expression for Roper excitation energy for a nucleon runs
as:[16]
\begin{equation}
\bigtriangleup E= \sqrt{\frac{K}{m r_{0}^{2}}}
\end{equation}
We have obtained the values as 0.757GeV and 0.699GeV in the free
space and in the normal nuclear density for the proton
respectively whereas for the neutron the values are obtained as
0.728GeV and 0.678GeV respectively.

\vskip .3in \noindent {\bf 4. Structure Function and Gottfried Sum
Rule}

 The study of the Structure function is important for the
understanding of the quark structure of the nucleons. The free
nucleon structure function in the non relativistic limit is
expressed as [19]:
\begin{equation}
F(x)=\frac{M}{8\pi^{2}}\int_{K_{min}}^{\infty}|\Psi(k)|^{2}dk^{2}
\end{equation}

 Where M is the mass of the nucleon and $\Psi(k)$ is the
normalized momentum space wave function and
$k_{min}$=M$|k-\frac{1}{3}|$. We have obtained the momentum space
wave function $\Psi(k)$ for the nucleon performing the Fourier
transform of the wave function in (12). The momentum space wave
function $\Psi(k)$ is obtained as [20]:
\begin{equation}
\Psi(k)= C_{1} k^{-1}j_{1}(kr_{0})
\end{equation}
where $C_{1}$ is the normalization constant and obtained as
2$\sqrt{3 \pi r_{0}}$.  We have estimated the structure function
of the proton $F_{2}^{p}$ and the neutron $F_{2}^{n}$ with the
input of $\Psi(k)$ in equation (19) using $r_{p}$ and $r_{n}$
estimated from the equations (15) and (16). The variation of the
ratio of the structure functions of the neutron to the proton with
the density of the nuclear medium have been displayed in the
Figure-4. The variation of the difference between the structure
functions of the proton and the neutron with the density of the
medium have been shown in the Figure-5.

The Gottfried Sum rule [21] $S_{G}$ can be expressed as,
\begin{equation}
S_{G}=\int_{0}^{1}\frac{F_{2}^{p}(x)-F_{2}^{n}(x)}{x}dx=\frac{1}{3}
\end{equation}
 We have estimated $S_{G}$ =
0.173 at $0<x<1$ for free nucleons. A  violation is observed. We
have investigated the variation of $S_{G}$ with
 $\rho$ and the variations is displayed
 in the Figure-6. Our
result agrees well with the existing theoretical and experimental
predictions. We have observed that the
 GT Sum increases with increasing
medium density. It rises up to 3.5 $\rho_{0}$ and then increases
rapidly till the critical density.

 \vskip .25in\vskip .3in \noindent {\bf 5. Result and Discussion}

In this Present work we have investigated the modifications of the
properties of the proton and the neutron in the nuclear medium.
The new thing in the current investigation is the inclusion of the
strange quark-anti quark contributions to the configuration of the
proton. We have computed the masses of the proton and the neutron
considering diquark as a fundamental constituent. The quasi
particle picture of the effective mass approximation for diquark
suggested by us [11] has been employed. The properties like
swelling, incompressibility, Roper resonance, structure function
and Gottfried Sum rule of nucleons have been studied with the
change of the nuclear matter density. The swelling of the proton
and the neutron in the nuclear medium have been shown in the
Figure-1. It has been observed that at the higher nuclear matter
density swelling of the proton is more the than the neutron which
indicates the fact that the effect is more pronounced in the
proton. We have observed 18 percent swelling in the proton radius
at critical density $\rho_{c}$ whereas neutron shows 13 percent
swelling. In the context of non topological model Wen et al [22]
have obtained 10-16 percent swelling of the nucleon radius at the
critical density. Noble et al [23] have estimated the increase in
size as 30 percent whereas Mathieu et al [16] have observed 25
percent swelling for the nucleon. We have obtained critical
density as 4.17$\rho_{0}$ and 4.69$\rho_{0}$ for the proton and
the neutron respectively which indicates that the neutron survives
more than the proton with respect to the nuclear medium. Mathieu
[16] have obtained 4.5$\rho_{0}$ whereas using the Global Colour
Symmetry model Chang et al [24] have estimated the critical
density as 8$\rho_{0}$ at which radius shows an infinite value
indicating a phase transition. Figure-2 represents the variation
of the masses with the nuclear density. It has been observed that
the nuclear medium causes a decrease in the mass of the nucleon.
We have observed a decrease of the masses of the proton form free
to the normal nuclear density as 0.8 percent whereas 0.6 percent
is observed for the neutron. The swelling of the proton is
estimated as 2.15 percent whereas for the neutron 1.96 percent has
been obtained at normal nuclear density $\rho_{0}$. Wen et al [22]
have also investigated the variation of masses of the nucleons
with medium density and have observed a decrease in mass. It is
interesting to observe here that the effect of the nuclear medium
is more pronounced in changing the size of the nucleon than the
mass of the nucleons from free to normal nuclear medium. Similar
observation is made by Wen et al [22]. Figure-3 represents the
variation of the incompressibility in the nuclear medium. The
decrease in K from free to normal nuclear medium are estimated as
$\Delta K$ 0.156 GeV for the proton and 0.132 GeV for the neutron.
Mathieu et al [16] have estimated the difference as 0.041 GeV.
Meissner et al [25] have investigated the incompressibility of the
nucleon and have obtained the value as 3GeV in the relativistic
approach. It is known that the incompressibility is related to the
excitation mode of the vibration and the Roper excitation energy
can be estimated form the knowledge of the incompressibility. We
have obtained the Roper resonance excitation energy as 0.757 GeV
for free proton and 0.699 GeV for the proton at normal nuclear
density whereas for the neutron we have obtained the values as
0.728 GeV and 0.678GeV respectively. The result shows that the
resonance state is more tightly bound in nuclear medium than in
the free nucleon. The current investigation yields comparatively
larger value of the Roper resonance compared to the experimental
value which is $\sim$ 500 MeV. Mathieu et al [16] have obtained
the values as 0.345 and 0.330 GeV respectively whereas Meissner et
al [25] have extracted it as 390 MeV. However it may be pointed
out that the radius parameter of the nucleon $r_{0}$ is not
exactly known and may shed some uncertainty in the results
particularly to the excitation energy of the Roper resonance.

The structure function $F_{2}(x,Q^{2}$) has been estimated and the
 variation with nuclear medium density $\rho $ has been studied. The
normalized wave function from (20) have been used to compute
$F_{2}(x)$. Using the radius $r_{p}$,$r_{n}$ from (15) and (16) we
have evaluated the structure functions $F_{2}^{n}$,$F_{2}^{p}$ for
x $\rightarrow$ 1. Results show some interesting behavior
regarding the ratio of structure functions and Gottfried Sum Rule
in medium. Figure-4 shows a gradual decrease of the ratio of the
structure functions of the neutron to the proton with increasing
$\rho$ up to 3.5 $\rho_{0}$ after which it falls sharply. The
variation in the difference between the structure functions of the
proton and the neutron i.e. ($F_{2}^{p}$ - $F_{2}^{n}$) have been
displayed in the Figure-5. It shows an increasing behavior as the
density of the nuclear medium increases. The variation of the
Gottfried Sum Rule with the nuclear medium density has been
studied and have displayed in Figure- 6. We have found $S_{G}$ to
be 0.173 at $0 < x < 1$ for the free nucleon. Experimental
prediction is $0.197 \pm 0.011 (st.) \pm 0.083 (sys.)$ at $0.02 <x
< 0.8$, 10 $<$ $Q^{2}<$90 GeV$^{2}$ [26] whereas NMC predicts the
value as 0.221 $\pm 0.008 (st.) \pm 0.019 (sys.)$ for $ 0.004 < x
< 0.8$ at $Q^{2}$ = 4 $GeV^{2}$ [27]. The result obtained in the
current investigation are found to be in the range of the
experimental predictions. Current investigation shows that the
values of the Gottfried Sum rule increases with the increase of
the medium density. It rises up to 3.5 $\rho_{0}$ and then
increases rapidly till the critical density. The variation of the
nuclear structure functions in the nuclear medium have been
studied by a number of authors [28]. Rozynek [5] has studied the
modification of the nuclear structure function in nuclear matter
above the saturation point in the relativistic mean field
approach. They have pointed out that the density evolution seems
to be stronger at the high densities. Cloet et al [29] have
studied the spin dependent structure functions of the nucleons in
the context of the modified NJL model. They have used the model to
study the modification of $F_{2}$ structure functions in the
nuclear medium and have suggested that the change the spin
structure function of a bound nucleon in the nuclear matter is
roughly twice as large as the change in the spin independent
structure function. It may be mentioned that recently Osipenko et
al [30] have reported the measurement of inclusive electron
scattering from CLAS at JLab. They have evaluated the $F_{2}$
structure functions and the corresponding moments. They have
speculated that special extension of nucleon changes in the
nuclear matter from the study of the structure function ratio of
carbon to deuteron.

In the current approach the proton is described as a five quark
system in the diquark-diquark-antiquark scheme incorporating the
strange contribution. This enable us to distinguish between the
proton and the neutron via their quark configurations.
Consequently their properties are distinguishable and show a
difference. The current approach is a naive one where the radius
of the relevant nucleon is the only parameter for estimating the
properties. The configurations of the proton and the neutrons with
diquark approach reproduce the results which are in reasonably
good agreement with other works. It may be mentioned that Trevisan
et al [31] have studied the structure function of a nucleon in a
statistical model incorporating strangeness content of proton.
They have studied the violation of the Gottfried sum rule and have
observed that NMC [27] results are well reproduced in this
approach.

The study of the structure functions and the violation of
Gottfried sum rule provides important information regarding the
structure of the nucleons.
 It has been suggested that the violation occurs due to the extrinsic
  contribution from the gluon
chains which are splitted into quark-antiquark pairs producing a
sea quark asymmetry. In the current work it has been suggested
that the 'intrinsic' sea has significant contribution and
reproduces the results well. Recently Chang et al[32] have
investigated the intrinsic charm quark contribution to proton via
uudc$\overline{c}$ configuration and investigated the five quark
Fock states. They have also investigated the light pentaquark
quark Fock states of the proton and have pointed out that the
'intrinsic' charm production is valance like with distribution
peaking at large x whereas extrinsic production is 'sea' like with
significant contribution at small x.  Results obtained are very
interesting and agrees with other available estimates. We have
observed that the violation increases with the density of the
medium indicating the fact that the quark density affects the
virtual quark sea of the proton and the neutron. In the present
investigation the effect is probed through the swelling of the
nucleon in medium. Studying the effect of the medium on the
modification of the properties of the nucleons are very important.
It throws some light on the quark structure of the nucleons and
their dynamics as the constituent quarks are attracted by the
neighbouring nucleons. We have tried to incorporate
$s\overline{s}$ effect in proton in current work. It may be
mentioned that the contribution due to strange quarks sea has been
incorporated as valence like whereas it has been suggested that
strange contribution is not more than 15 to 20 percent to the
magnetic moment. However it may be mentioned that there is a
number of works [33-34] where the strange contribution has been
treated as valence contribution and the proton is described as a
penta quark system particularly to study the magnetic moments of
the proton. It is pertinent to point out here that the proton is a
complicated object with the contributions from the valance quarks,
the gluons and the strange quark sea. It is a challenging job to
reveal the proton structure. Only 50 percent of the proton
momentum is carried by the valence quarks [33]. It has also been
suggested that 30 percent of the proton spin is carried by strange
sea. The current experimental results hints to the fact that the
proton flavour structure may not be limited to u and d only.
Probing the structure of the proton is a challenging job to the
modern day particle physics. It is interesting to study the effect
of the strange contribution to the properties of the proton.We
have incorporated the strange pair contribution to the proton
configuration as five quark system and have investigated the
possibility whether the sea quark asymmetry between the two
nucleons can be attributed to the strange sea quark contribution
to the proton flavour structure. It may be asserted that the quark
structure and the medium play an important role in modifying the
properties of the nucleons. More investigations on the the
structure of the proton and its resonance states in medium with
vector diquark would be studied in our future works.

\vskip .25in

{\bf Acknowledgement} .Authors are Thankful to University Grants
Commission (UGC), New Delhi,India for financial assistance ( F.No.
37-217/2009(SR) .

\pagebreak {\bf References}:-

\vskip .25in

\noindent 1. European Muon Collaboration Aubert J. J. et al.: The
ratio of the nucleon structure functions $ F_{2}$(N) for iron and
deuterium. Phys. Lett.{\bf B 123}, 275 (1983);

Ashman J.et al.: A measurement of the spin asymmetry and
determination of the structure function g1 in deep inelastic
muon-proton scattering. Phys. Lett.{\bf B 206}, 364 (1988)

\noindent 2. Seely J. : New measurement of european collaboration
effect in very light nuclei. Phys.Rev.Lett. {\bf 103}, 202301
(2009)

\noindent 3.  Wu W. ,Hong S.: Nucleon properties in the nuclear
medium. Chinese Phys.C {\bf 32}, Suppl.II 81 (2008)

\noindent 4. Kim K. S.,Cheoun M.K.,Cheon I.-T.,Chung Y.: Effect of
the form factor on exclusive(e,e'p) reactions. Euro.Phys.J {\bf
8A}, 131 (2000)

\noindent 5. Rozynek J.: Finite pressure corrections to nucleon
structure function inside a nuclear medium. Acta Physics Polonica
{\bf B 5} no-2 (2012)

\noindent 6.  Higinbotham D. W.,Gemez J.,Piasetzky E.: Nuclear
scaling and the EMC effect. Bulletin of the American Physical
Society 2010 Fall meeting of the APS Division of Nuclear Physics
{\bf 55} no-14 (2010)

\noindent 7. Alekseev M.G. et al. (COMPASS Collab.): The
Spin-dependent Structure Function of the Proton $g_1^p$ and a Test
of the Bjorken Sum Rule. Phys.Lett. {\bf B 690} 466 (2010);

 Airapetian A. et al (HERMES Collab.): Precise determination of the spin
 structure function g1 of the proton, deuteron, and neutron.  Phys.Rev. {\bf D 75} 012007 (2007);

  Ito T. M.et al. (SAMPLE Collaboration): Parity-Violating electron deuteron
   scattering and the proton's neutral weak axial vector form factor. Phys.Rev.Lett. {\bf 92} 102003 (2004 );

  Armstrong D. S. et al.(Go Collaboration): Strange-quark contributions to parity-violating
  asymmetries in the forward G0 electron-proton scattering experiment.
  Phys. Rev. Lett. {\bf 95} 092001 (2005)

\noindent 8. Zou B. S., Riska D. O.: The $s\bar s$ component of
the proton and the strangeness magnetic moment. Phys. Rev.
Lett.{\bf 95}, 072001 (2005)

\noindent 9. Riska D. O.,Zou B. S.: The strangeness form factors
of the proton.Phys. Lett. {\bf B 636}, 265 (2006)
arXiv:0512102[Nucl-th](2005)

\noindent 10. An C. S., Riska D.O., Zou  B.S.: Strangeness spin,
magnetic moment and strangeness configurations of the proton.
Phys.Rev. {\bf C 73}, 035207 (2006)

\noindent 11. Bhattacharya A.,Chandra A.Chakrabarti B.,Sagari
A.:The heavy-light baryon masses in quasi-particle approach.
Eur.Phys.J.Plus {\bf126}, 57 (2011)

 \noindent 12. Haug A.: Theoretical Solid State Physics
(Pergamon Press Oxford) p-100 (1975)

\noindent 13. Lucha W., Sch$\ddot{o}$berl F. F., Gromes D.: Bound
states of quarks. Phys. Rep. {\bf 200}, 127 (1991)

 \noindent 14. Maris P.: Electromagnetic properties of diquarks.Few Body Syst. {\bf 35}, 117 (2004)

\noindent 15. Karliner M. and Lipkin H. J.: A diquark–triquark
model for the KN pentaquark. Phys.Lett. {\bf B 575}, 249 (2003)

\noindent 16. Mathiue P., Watson J. S.: A flux-tube model of
hadrons in nuclear matter. Canadian J. of Phys. {\bf 64}, 1389 (
1986)

 \noindent 17. Bhattacharya A.,Banerjee S.N.,Chakrabarti
 B.,Banerjee S.: A study on the structure of proton. Nucl.Phys.B {\bf  142}, 13 (2005)

\noindent 18. Chen B.,Wang D.X.,Zhang A.: $J^{P}$ assignment of
$\Lambda_{c}^{+}$ baryons. Chin. Phys. {\bf C 33}, 1327 (2009)

\noindent 19.  R$\acute{u}$jula A. De., F.Martin: Structure of
hadron structure functions. Phys. Rev. {\bf D 22 }, 1787 ( 1980)

 \noindent 20. Bhattacharya A.,Remadevi T.L.,Banerjee S.N.: On the studies of the structure function of nucleon. Fizika {\bf B
4}, 59 (1995)

\noindent 21. Gottfried K.: Sum Rule for High-Energy
Electron-Proton Scattering. Phys.Rev.Lett. {\bf 18}, 1174 (1967)

 \noindent 22. Wen W., Shen H.: Modification of nucleon properties in nuclear matter and finite nuclei.
  Phys.Rev.{\bf C 77}, 065204
 (2008)

 \noindent 23. Noble J. V.: Modification of the Nucleon's Properties in Nuclear Matter.
   Phys.Rev.lett.{\bf B 46}, 412 (1981)

 \noindent 24. Chang L., Liu Y. X., Guo H. : Density dependence
  of nucleon radius and mass in the global color symmetry model of QCD with a sophisticated effective gluon
  propagator. Nucl. Phys.{\bf A 750}, 324 (2005)

 \noindent 25. Meissner T.,Grummer F.,Goeke K.:Roper resonances and generator co-ordinate method in
 the chiral soliton model.
 Phys.Rev.{\bf D 39}, 1903 (1989)

 \noindent 26. Aubert J. J. et al: Measurements of the nucleon structure functions
 $F_{2}^{N}$ in deep inelastic muon scattering from deuterium and comparison
with those from hydrogen and iron.  (EMC Collab.)Nucl.Phys.{\bf B
 293}, 740 (1987)

\noindent 27. Arneodo M. et al., (New Muon Collab.): A
Reevaluation of the Gottfried Sum. Phys.Rev.{\bf D 50}, 1 (1994)

 \noindent 28. Villate J. E. and Koltun D. S.: Nuclear medium modification of nucleons via a Bethe ansatz model.
 Phys.Rev.{\bf C 43}, 1967 (1991)

  Athar M. S.:Nuclear medium modification of the $F_{2}(x,Q^{2})$ structure
function. arXiv:0910.4879[nucl-ph] (2011);

 Smith  J. R., Miller G. A.: Return of the EMC effect: Finite nuclei. Phys.Rev.{\bf C 65}, 055206 (2002)

 \noindent 29. $Clo\ddot{e}t$ I. C.,Bentz W.,Thomas A.W.: Spin-dependent structure functions
  in nuclear matter and the polarized EMC effect.  Phys.Rev.Lett. {\bf 95}
 052302 (2005)

 \noindent 30. Osipenko M. et al: Measurement of the
 Nucleon Structure Function F2 in the Nuclear Medium and Evaluation of its Moments. Nucl.Phys.{\bf A 845} 1 ( 2010)

 \noindent 31. Trevisan L. A.,Frederico T.,Tomio L. : Strangeness content and structure function
 of the nucleon in a statistical quark model. Eur.Phys.J {\bf C 11} 351 (1999)

 \noindent 32. Chang W. C.,Peng J.C.: Extraction of Various Five-Quark Components of the Nucleons. Phys.Rev.Lett.{\bf 106}
 252002 (2011)

\noindent 33. Bijker R.: Strange form factor of the proton in a
two-component model. J. Phys. G:Nucl. Part. Phys. {\bf 32}, L49
(2006)

\noindent 34. Zou B.S.: Five components in baryons. arXiv:
1001.1084[nucl-th] (2010)

\newpage
\vskip 0.25 in {\bf Figure Captions}:

Fig.1.The ratio of the nucleon radius in nuclear matter to that in
free space(r/$r^{0}$) as a function of nuclear matter
density($\rho$).
 \vskip 0.25 in
 Fig.2 The ratio of the nucleon
mass in nuclear matter to that in free space({$M^{*}$/M})as a
function of nuclear matter density($\rho$) \vskip 0.25 in

Fig.3.The ratio of Incompressibility of nucleon in nuclear matter
to that in free space (K/$K_{0}$) as a function of nuclear matter
density($\rho$).

\vskip 0.25in
 Fig.4The ratio of the structure function of neutron
to that of proton as a function of nuclear matter density($\rho$)

\vskip 0.25 in
 Fig.5.The difference between the Structure function
of proton and neutron as a function of nuclear matter
density($\rho$)

\vskip 0.25in
 Fig.6.The variation of GT Sum Rule  as a function of nuclear
 matter density($\rho$).
 \newpage

\end{document}